# Small Signal Analysis of SQUID Direct Readout Schemes

Yongliang Wang [a,b,c,*] , Guofeng Zhang [a,b], Shulin Zhang [a,b] ,

Liangliang Rong [a,b], Yong Wang [c], Xiaoming Xie [a,b]

[a] *State Key Laboratory of Functional Materials for Informatics, Shanghai Institute of Microsystem and Information Technology (SIMIT), Chinese Academy of Sciences (CAS), Shanghai 200050, China*
[b] *Chinese Academy of Science, Center for Excellence in Superconducting Electronics, Shanghai 200050, China*
[c] *Department of Automation, University of Science and Technology of China, Hefei 230027, China*

[*]Corresponding author. Tel.: +86 02162511070; Fax: +86 02162127493.

E-mail address: wangyl@mail.sim.ac.cn

## Abstract

  To better understand working principles of Superconducting Quantum Interference Device (SQUID) direct readout schemes which are working in different bias and amplifier modes with different internal feedback schemes, we present the complete circuit analyses based on SQUID small signal model. SQUID bias and amplifier circuits are analyzed using SQUID Thevenin's equivalent circuit, and the general equivalent circuit of SQUID with different internal feedback schemes is derived and analyzed with trans-impedance amplifier model. Transfer characteristics and noise performances of different direct readout schemes are analyzed and experimentally characterized. It is shown that, amplifier noise suppression is only depended on SQUID flux-to-voltage transfer coefficient and irrelevant to the configuration of bias and amplifier; SQUID with internal feedback scheme improves the transfer coefficient with voltage feedback, and regulates the dynamic resistance with current feedback.

  *Keywords:* SQUID, direct readout, voltage feedback, current feedback.

# 1. Introduction

Direct Current Superconducting Quantum Interference Device (DC SQUID) is sensitive non-linear flux-to-voltage convertor [1], which has to be used in the Flux-Locked Loop (FLL) to realize linear flux measurement [2]. Typical FLL consists of SQUID, amplifier, integrator and flux feedback circuit as shown in Fig. 1. Readout circuit is referred as SQUID and its amplifier in this paper. Room temperature amplifier is directly coupled with SQUID without Flux Modulation (FM) scheme [3] in the direct readout schemes, which were developed rapidly to build simple and compact readout electronics for practical multichannel SQUID systems, e.g., Magneto-Cardiogram (MCG). The challenge of direct readout FLL is the noise mismatching between SQUID and amplifier which voltage noise is usually ten times of SQUID intrinsic noise and dominates the noise performance of FLL.

Several internal feedback circuits around SQUID were introduced to suppress amplifier noise. Additional Positive Feedback (APF) was firstly presented by D.Drung *et al.* in early 1990s [4-6], and the same scheme working under voltage bias was called Noise Cancellation (NC) [7, 8]. Furthermore, Bias Current Feedback (BCF) circuit was introduced to suppress external current noise in addition to APF [9, 10], meanwhile, the SQUID Bootstrap Circuit (SBC) evolved from APF and BCF was also developed for working under voltage bias [11, 12]. Overall behaviors and working principles of those readout schemes were usually analyzed by considering SQUID and amplifier separately [13], and studied individually [14, 15]. However, those various readout concepts are confusing users in their readout electronics design. The more common circuit analyses of SQUID direct readout schemes are required.

In this paper, we will present complete circuit analyses of SQUID direct readout schemes based on the small signal circuit model in form of Thevenin's equivalent circuit. We will firstly present two general configurations of bias and amplifier, and bring out the general equivalent circuit of all the SQUID internal feedback schemes, and then analysis the transfer coefficient and noise performance of low noise direct readout schemes which are the combinations of different bias modes and internal feedback schemes. Finally, the experimental results are presented and discussed.

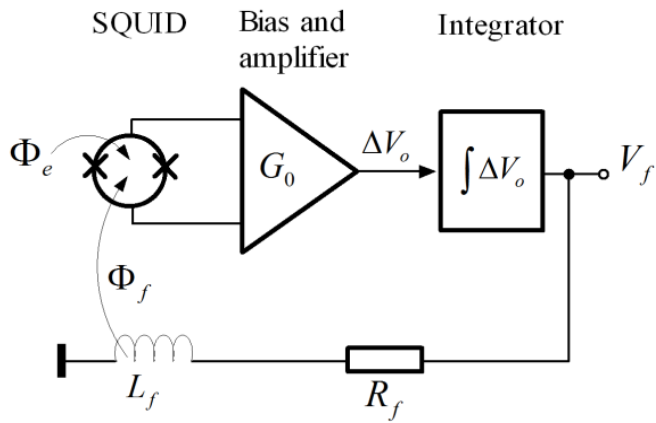

Fig. 1. Equivalent circuit of direct readout FLL.

## 2. SQUID bias and amplifier circuit

*2.1 Two SQUID bias and amplifier circuits*

SQUID is directly coupled with the bias and amplifier circuit in the direct readout FLL as shown in Fig.1. There are only two common configurations of bias and amplifier circuit in the existing SQUID direct readout schemes [2, 16]. The one shown in Fig. 2a is called Current-Bias-Voltage-Amplifier (CBVA) mode, the other shown in Fig. 2c is called Voltage-Bias-Current-Amplifier (VBCA) mode. The details are in the following:

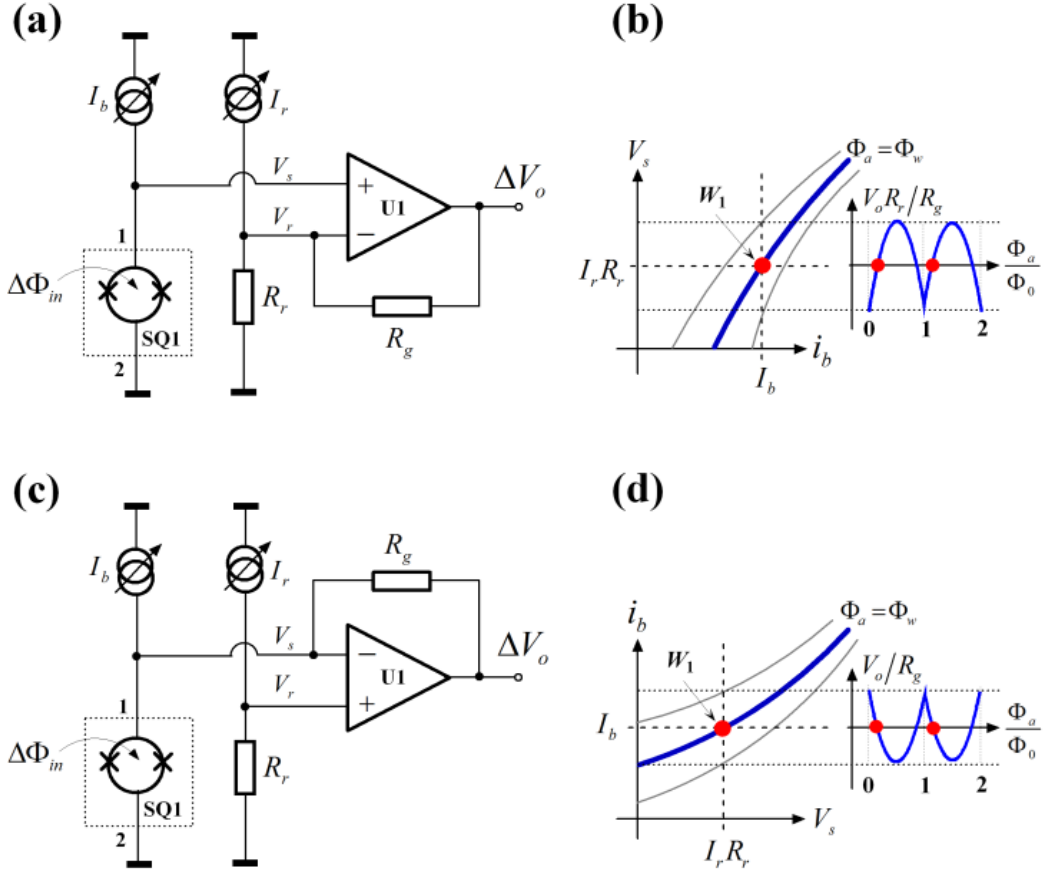

Fig. 2. (a) SQUID direct readout circuit with CBVA, (b) Illustration of flux-to-voltage characteristics read out with CBVA, (c) SQUID direct readout circuit with VBCA, (d) Illustration of flux-to-current characteristics read out with VBCA.

1) Bias circuit is an H-type bridge, in which two-terminal SQUID SQ1 and reference resistor $R_r$ are two compatible arms. Both are biased with current $I_b$ and $I_r$ respectively, where, $V_s$ is voltage output of SQ1, and $V_r$ is voltage of reference resistor $R_r$. $V_s$ can be described with a function of bias current $i_b$ and applied flux $\Phi_a$, i.e., $V_s = f(i_b, \Phi_a)$, according to its bias-dependant flux-to-voltage transfer characteristics.

2) Differential amplifier is implemented using a low noise operational amplifier in closed-loop with feedback resistor $R_g$ connected at its non-inverting input and voltage output as shown in Fig. 2a and c. In practical, $R_g \gg R_r$, the current shunted by $R_g$ is neglected in the following analyses.

3) In CBVA mode, $V_s$ is connected to the non-inverting input, and $V_r$ is connected to the inverting input of amplifier, where, SQ1 is biased under current $I_b$, and $R_r$ is driven by current from $I_r$ and $V_o$, thus, $V_s = f(I_b, \Phi_a)$, $V_r = R_r(I_r + V_o / R_g)$.

4) In VBCA mode, the connection is reversed, where, SQ1 was driven by current from $I_b$ and $V_o$, and $R_r$ is biased by $I_r$, thus, $V_s = f((I_b + V_o / R_g), \Phi_a)$, $V_r = R_r I_r$.

Since, differential inputs of closed-loop operational amplifier are "shorted" [17], i.e., $V_s = V_r$. The $\Phi_a$-to-$V_o$ characteristics read out with CBVA exhibit the flux-to-voltage characteristics of SQ1 as illustrated in Fig. 2b. The $\Phi_a$-to-$V_o$ characteristics read out with VBCA exhibit the flux-to-current characteristics of SQ1 as illustrated in Fig. 2d. Characteristics of SQUID read out with CBVA and VBCA are totally different [11].

However, when FLL is working in closed-loop with null input of integrator [2], i.e., $V_o=0$, working point of SQ1 in both amplifier modes is same as defined in (1), where, current of SQ1 is $I_b$, and voltage output is $I_r R_r$; $\Phi_w$ is the background flux on the working point in the absence of external flux.

$$f(I_b, \Phi_w) = I_r R_r \tag{1}$$

*2.2 Small signal analysis of amplifier circuits*

Around the working point, small signal model of SQ1 is extracted from its total differential of SQUID flux-to-voltage function as:

$$\begin{aligned}\Delta V_s &= f(I_b + \Delta I_b, \Phi_w + \Delta \Phi_a) - f(I_b, \Phi_w) \\ &\cong \frac{\partial V_s}{\partial \Phi_a}\Delta \Phi_a + R_d \Delta I_b\end{aligned} \tag{2}$$

Where, $\partial V_s/\partial \Phi_a$ is the flux-to-voltage transfer coefficient, and $R_d$ is the dynamic resistance, they are defined as:

$$\begin{cases}\dfrac{\partial V_s}{\partial \Phi_a} = \dfrac{\partial f}{\partial \Phi_a}\bigg|_{\substack{i_b=I_b \\ \Phi_a=\Phi_w}} \\ R_d = \dfrac{\partial f}{\partial i_b}\bigg|_{\substack{i_b=I_b \\ \Phi_a=\Phi_w}}\end{cases} \tag{3}$$

SQUID small signal model is implemented in form of the Thevenin's equivalent circuit [18], which is a flux-driven voltage source in series with an internal resistor as shown in Fig. 3a and b. $(\partial V_s/\partial \Phi_a)\Delta \Phi_a$ is thus the flux-driven voltage, and $R_d$ is the internal resistance.

SQUID working with CBVA and VBCA implement small signal flux-to-voltage conversion before integrator with $\Delta \Phi_{in}$ as flux input and $\Delta V_o$ as voltage output. $\Delta V_o/\Delta \Phi_{in}$ is thus the overall flux-to-voltage transfer coefficient of direct readout circuit, and is part of the open-loop gain of FLL.

The equivalent small signal readout circuits are shown in Fig. 3a and b, where, $\Delta \Phi_{in}$ comes from external input flux $\Phi_e$ and feedback flux $\Phi_f$ as shown in Fig. 1. Since, $\Delta \Phi_a = \Delta \Phi_{in}$, the small signal voltage of SQ1 is $(\partial V_s/\partial \Phi_a)\Delta \Phi_{in}$. SQ1 with CBVA is known as non-inverting amplifier circuit with gain of $R_g/R_r$ [17]. Its $\Delta \Phi_{in}$-to-$\Delta V_o$

transfer coefficient is derived as:

$$\left(\frac{\Delta V_o}{\Delta \Phi_{in}}\right)_{CBVA} = \frac{R_g}{R_r} \frac{\partial V_s}{\partial \Phi_a} \qquad (4)$$

Meanwhile, SQ1 with VBCA is known as inverting amplifier circuit with gain of $-R_g/R_d$ [17]. Its $\Delta\Phi_{in}$-to-$\Delta V_o$ transfer coefficient is written as:

$$\left(\frac{\Delta V_o}{\Delta \Phi_{in}}\right)_{VBCA} = -\frac{R_g}{R_d} \frac{\partial V_s}{\partial \Phi_a} \qquad (5)$$

If $R_r = R_d$, the overall transfer coefficients in (4) and (5) are equivalent with opposite sign. However, $\partial V_s/\partial \Phi_a$ can be directly characterized in CBVA mode, but not in VBCA mode.

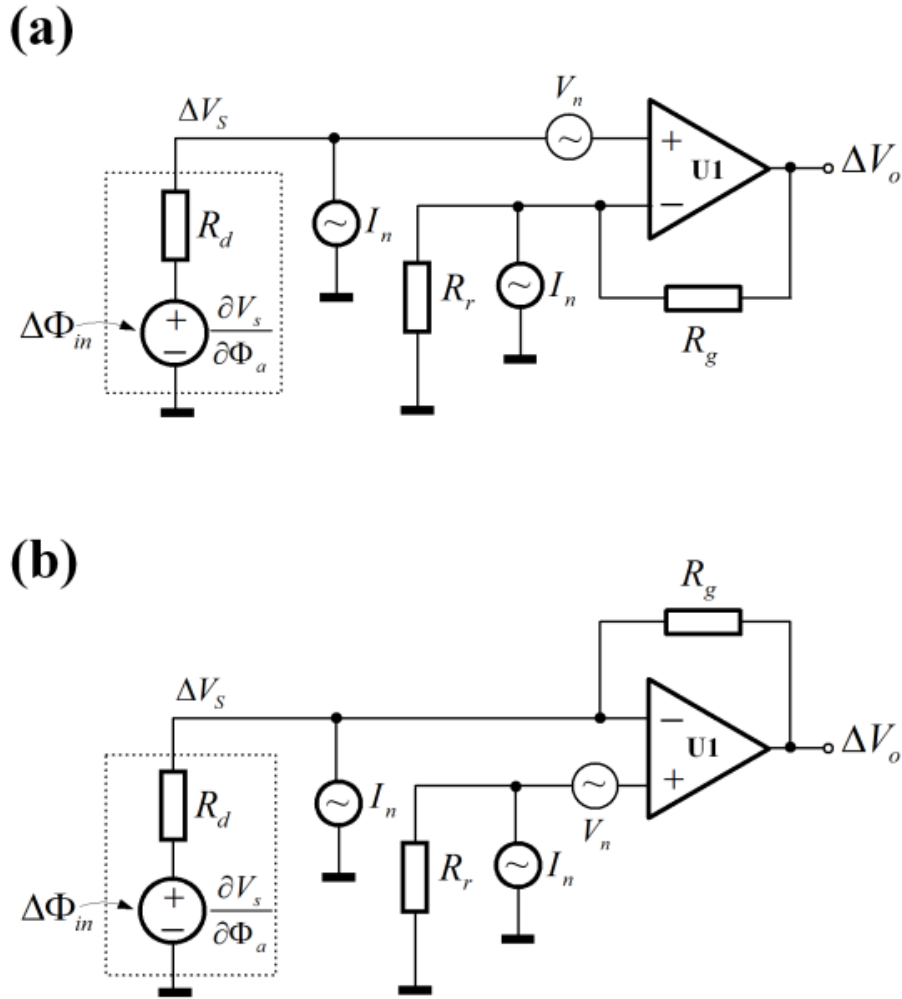

Fig. 3. (a) Small signal circuit of SQUID working with CBVA, (b) Small signal circuit of SQUID working with VBCA.

Equivalent flux noise of amplifier $\Phi_n$ is the another key parameter which determines the noise performance of FLL. By introducing voltage noise $V_n$ and current noise $I_n$ of operational amplifier to the equivalent small signal readout circuits, the total noise contributions in both VBCA and CBVA circuits are the Root Mean

Square (RMS) of $V_n$ and voltage noises generated by $I_n$ on $R_d$ and $R_r$. $\Phi_n$ is derived as:

$$\Phi_n = \frac{\sqrt{V_n^2 + (I_n R_d)^2 + (I_n R_r)^2}}{\partial V_s / \partial \Phi_a} \cong \frac{V_n}{\partial V_s / \partial \Phi_a} \quad (6)$$

Voltage noise $V_n$ is the dominant noise, and current noise $I_n$ can be neglected if $R_d$ and $R_r$ meet the noise impedance matching condition as expressed in (7), where, $R_n$ is defined as the noise impedance of amplifier U1.

$$R_d < R_n; R_r < R_n; R_n = \frac{V_n}{I_n} \quad (7)$$

It is shown that equivalent flux noise of amplifier is irrelevant to amplifier configurations [13, 19], and is only determined by $\partial V_s/\partial \Phi_a$ of SQUID as noise impedance matching condition is satisfied.

## 3. SQUID with internal feedback circuit

### 3.1 SQUID internal feedback schemes

SQUID with internal feedback circuits are developed to improve transfer coefficient for amplifier noise suppression. There are three typical internal feedback schemes as shown in Fig. 4.

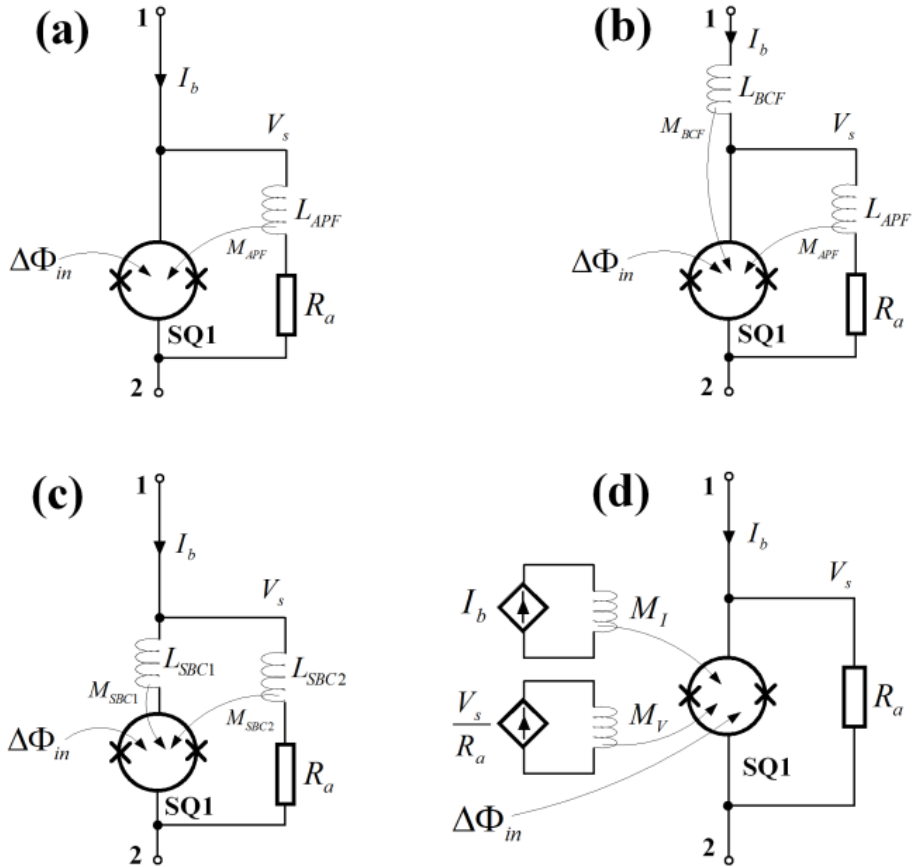

Fig. 4. (a) The APF scheme, (b) The BCF + APF scheme, (c) The SBC scheme, and (d) the general equivalent circuit of internal feedback schemes.

1) The APF scheme is shown in Fig. 4a, in which, SQUID is shunted by a resistor $R_a$ in series with inductance $L_{APF}$. $L_{APF}$ is coupled to SQ1 with mutual inductance $M_{APF}$. The flux internally coupled to SQ1 by shunt current through $L_{APF}$ is $M_{APF}V_s / R_a$.

2) Fig. 4b shows the BCF+APF scheme which is the BCF circuit in addition to APF scheme. BCF creates a flux feedback driven by bias current $I_b$ through inductance $L_{BCF}$, which is coupled to SQ1 with mutual inductance $M_{BCF}$. The fluxes internally coupled to SQ1 through $L_{APF}$ and $L_{BCF}$ are written as $M_{APF}V_s / R_a - M_{BCF}I_b$. Here, the minus sign indicates the fluxes generated by bias current $I_b$ and shunt current $V_s / R_a$ are opposite.

3) The SBC scheme is also on the basis of APF scheme as shown in Fig. 4c. The inductance used for APF is renamed as $L_{SBC2}$, and an extra inductance $L_{SBC1}$ is inserted in series with SQ1 which current is $I_b - V_s / R_a$. $L_{SBC1}$ and $L_{SBC2}$ are coupled to SQ1 with mutual inductance $M_{SBC1}$ and $M_{SBC2}$ respectively. If fluxes generated by current through SQ1 and current through $R_a$ are assumed opposite, the total fluxes internally coupled to SQ1 are rewritten as $(M_{SBC1} + M_{SBC2}) V_s / R_a - M_{SBC1} I_b$.

It shows that there are two internal flux feedback modes in those internal feedback schemes. The one driven by $I_b$ is defined as Current Feedback (CFB). The other one driven by $V_s$ is defined as Voltage Feedback (VFB). Their mutual inductances coupled with SQ1 are defined as $M_I$ and $M_V$. The configurations of $M_I$ and $M_V$ in the schemes shown in Fig. 4 are summarized in Table 1.

Table 1. Configurations of $M_V$ and $M_I$ in different internal feedback schemes.

| Scheme | Value of $M_V$ | Value of $M_I$ |
|---|---|---|
| APF | $M_V = M_{APF}$ | $M_I = 0$ |
| BCF+APF | $M_V = M_{APF}$ | $M_I = M_{BCF}$ |
| SBC | $M_V = M_{SBC1} + M_{SBC2}$ | $M_I = M_{SBC1}$ |

Therefore, the internal feedback schemes are equivalent if they are configured with the same $M_V$ and $M_I$, e.g., the BCF+APF scheme and SBC scheme are equivalent, if $M_{APF} = M_{SBC1} + M_{SBC2}$, and $M_{BCF} = M_{SBC1}$.

All the schemes above are unified with a general equivalent circuit as shown in Fig. 4d. The CFB and VFB circuits are functioned as two flux generators, which one is driven by bias current $I_b$, and the other one is driven by shunt current $V_s/R_a$; they are coupled to SQUID with mutual inductance $M_I$ and $M_V$ respectively. Here, the impedance of inductances is negligible for frequency limit of circuits.

*3.2 Small signal analysis of internal feedback circuit*

By using SQUID small signal model in the general equivalent circuit, we can have the equivalent small signal circuit of SQUID internal feedback schemes as shown in Fig. 5a.

The current-driven flux generators of CFB and VFB cooperated with small signal SQUID are functioned as two Trans-Impedance Amplifiers (TIAs) [17], which covert current $\Delta I_b$ or $\Delta V_s/R_a$ into small voltages. The trans-impedances of TIA are defined with mutual inductances and flux-to-voltage transfer coefficient as:

$$\begin{cases} R_I = \dfrac{\partial V_s}{\partial \Phi_a} M_I \\ R_V = \dfrac{\partial V_s}{\partial \Phi_a} M_V \end{cases} \tag{8}$$

With TIA model and concepts of trans-impedances $R_I$ and $R_V$, circuit in Fig. 5a is internalized into linear circuit as shown in Fig. 5b, in which, SQUID small signal voltage consists of three components, which one is voltage source $(\partial V_s/\partial \Phi_a)\Delta\Phi_{in}$, and the other two are dependant voltage sources $R_I\Delta I_b$ and $R_V\Delta V_s/R_a$.

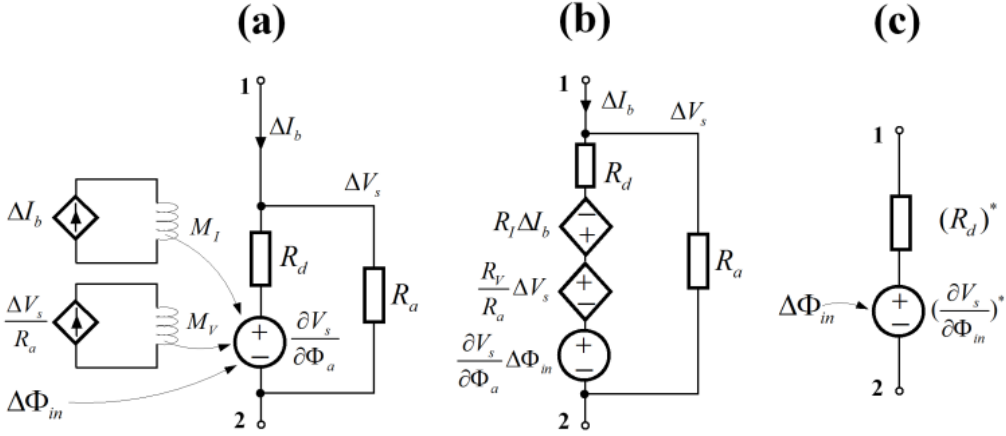

Fig. 5. (a) The equivalent small signal circuit of SQUID internal feedback schemes, (b) The simplified equivalent circuit with TIA model, (c) The unified Thevenin's equivalent circuit of SQUID internal feedback schemes.

Fig. 5c shows the unified Thevenin's equivalent circuit of all the SQUID with internal feedback schemes derived from linear electric circuit in Fig. 5b, where, $(\partial V_s/\partial \Phi_{in})^*$ is the equivalent transfer coefficient and $(R_d)^*$ is the equivalent dynamic resistance, they are written as:

$$(\frac{\partial V_s}{\partial \Phi_{in}})^* = \frac{\partial V_s}{\partial \Phi_a} \cdot \frac{1}{1-(R_V-R_d)/R_a} \tag{9}$$

$$(R_d)^* = R_d \cdot \frac{1-R_I/R_d}{1-(R_V-R_d)/R_a} \tag{10}$$

It is shown that the flux-to-voltage transfer coefficient $(\partial V_s/\partial \Phi_{in})^*$ is only improved by VFB with trans-impedance $R_V$; while, the dynamic resistance $(R_d)^*$ is increased by VFB with $R_V$, and decreased by CFB with $R_I$ [12, 13].

The numerators and denominators in (9) and (10) should always be positive to keep circuit stable, thus, the critical conditions of $R_I$ and $R_V$ are derived as:

$$R_I < R_d; R_V < R_d + R_a \tag{11}$$

By using TIA model, SQUID internal feedback schemes are turned into conventional electric circuits, which are easily analyzed with fundamental electric circuit theorems.

# 4. Low noise direct readout schemes

*4.1 Configurations of low noise direct readout schemes*

All the low noise direct readout schemes being reported, e.g., APF, NC, BCF+APF, and SBC, are the applications of internal feedback schemes working in CBVA or VBCA mode as shown in Fig. 6. APF and BCF+APF schemes are read out with CBVA, while, NC and SBC schemes are read out with VBCA.

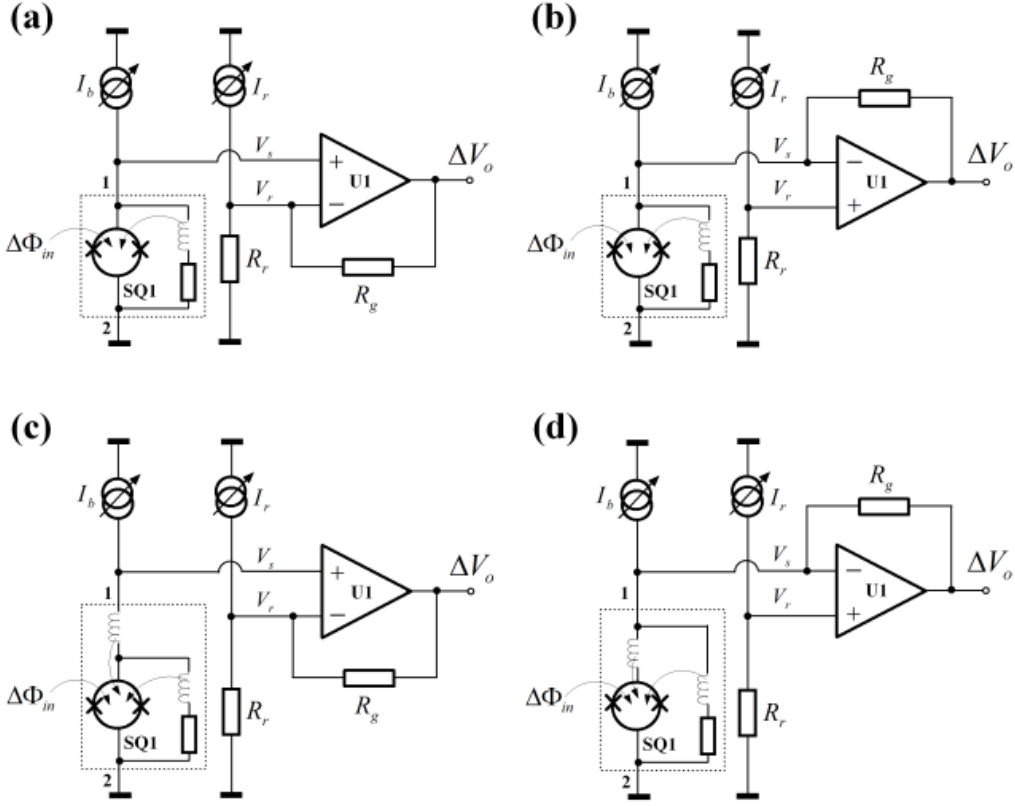

Fig. 6 (a) The APF direct readout circuit, (b) The NC direct readout circuit, (c) The BCF + APF direct readout circuit, and (d) the SBC direct readout circuit.

*4.2 Performances improvement*

By using the unified Thevenin's equivalent circuit shown in Fig. 5c, we can have the small signal readout circuit of low noise readout schemes as shown Fig. 7a and b. Here, the current noises of preamplifier are neglected if the noise impedance matching condition in (7) is satisfied.

After using internal feedback scheme, the overall amplified transfer coefficients of SQUID direct readout circuits are derived as:

$$(\frac{\Delta V_o}{\Delta \Phi_{in}})^*_{CBVA} = \frac{R_g}{R_r}(\frac{\partial V_s}{\partial \Phi_{in}})^*$$

$$= (\frac{\Delta V_o}{\Delta \Phi_{in}})_{CBVA} \frac{1}{1-(R_V - R_d)/R_a} \quad (12)$$

$$\left(\frac{\Delta V_o}{\Delta \Phi_{in}}\right)^*_{VBCA} = -\frac{R_g}{(R_d)^*}\left(\frac{\partial V_s}{\partial \Phi_{in}}\right)^* \quad (13)$$
$$= \left(\frac{\Delta V_o}{\Delta \Phi_{in}}\right)_{VBCA} \frac{1}{1 - R_I/R_d}$$

Compared with the transfer coefficient definitions in (4) and (5), the characteristics read out with CBVA are modified by VFB with trans-impedance $R_V$, and the characteristics read out with VBCA are modified by CFB with trans-impedance $R_I$ [10, 11].

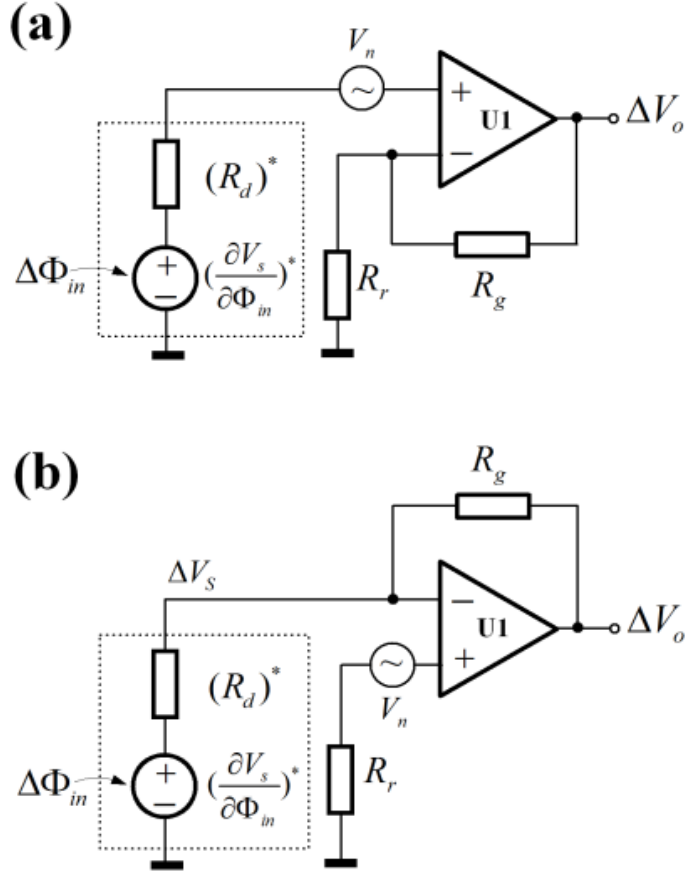

Fig. 7. (a) The small signal equivalent circuit of APF and APF+BCF readout schemes, and (b) the small signal equivalent circuit of NC and SBC readout schemes.

Meanwhile, the equivalent flux noise of amplifier after using internal feedback circuit is rewritten as:

$$(\Phi_n)^* = \frac{V_n}{(\partial V_s/\partial \Phi_{in})^*} \quad (14)$$
$$= \Phi_n \cdot (1 - (R_V - R_d)/R_a)$$

Here, $\Phi_n$ is the flux noise of preamplifier working with SQUID without internal feedback as defined in (6). Thus, $(1-(R_V-R_d)/R_a)$ is the noise suppression factor after using internal feedback scheme, and is only decreased by VFB with $R_V$.

CFB circuit in the readout is only used to regulate dynamic resistance $(R_d)^*$ for noise impedance matching in the applications where current noise is significant [20].

## 5. Experimental results

The experimental set up for validation of SQUID low noise readout schemes is shown in Fig. 8. The bias and amplifier circuit can be switched between CBVA and VBCA modes by reversing the connections of bias bridge and differential amplifier with a Double-Pole-Double-Throw (DPDT) switch SW1. AD797 (from Analog Device) with white voltage noise of 1nV/√Hz (at 1k Hz) was used as amplifier directly coupled with SQUID. Its noise impedance $R_n$ is about 500 Ω. Resistor $R_p$ is matching with $R_g$ to balance the input impedances of amplifier, $R_p = R_g$.

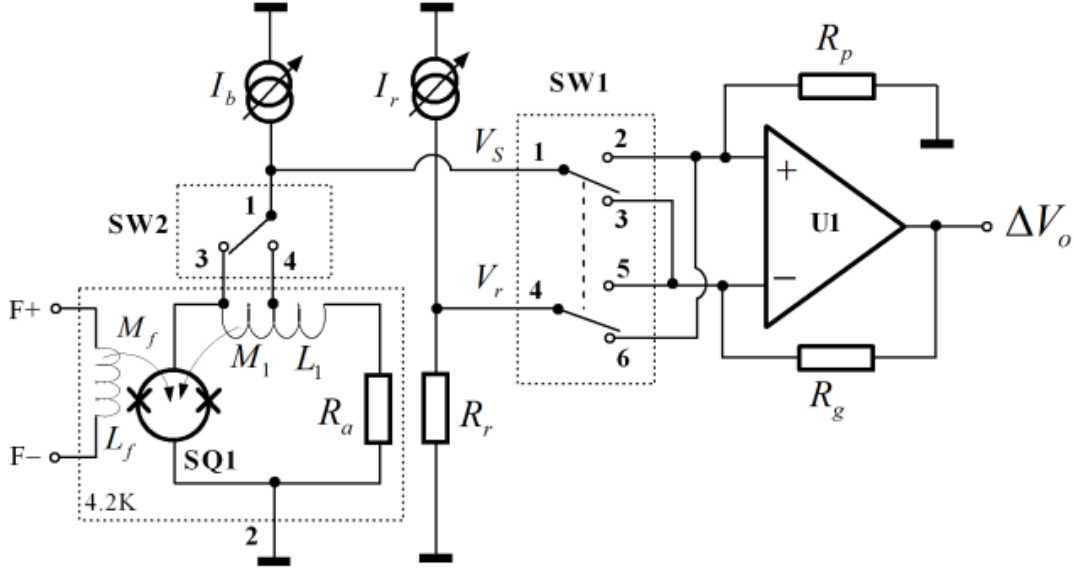

Fig. 8. SQUID direct readout circuit configurable for testing.

Table 2. Parameters of SQUID and readout circuit.

| Parameter | Symbol | Value (Unit) |
|---|---|---|
| Bias current of SQ1 | $I_b$ | ≈ 20 μA |
| Voltage swing of SQ1 | $V_{s\_pp}$ | ≈ 40 μV |
| Loop inductance of SQ1 | $L_s$ | ≈ 60 pH |
| Input coil coupling | $1/M_1$ | ≈ 5.8 μA/$\Phi_0$ |
| Feedback coil coupling | $1/M_f$ | ≈ 56 μA/$\Phi_0$ |
| Resistor | $R_a$ | 27 Ω |
| Resistor | $R_r$ | 3 Ω |
| Resistor | $R_g, R_p$ | 2 kΩ |
| FLL feedback resistor | $R_f$ | 10 kΩ |

A low-Tc niobium-based dc SQUID integrated with a feedback coil $L_f$ and an input coil $L_1$ was applied to build a SQUID internal feedback circuit. A tap was drawn out from $L_1$ as the third terminal. Two terminals of $L_1$ were connected to a Single-Pole-Double-Throw (SPDT) switch SW2 (terminal 3 and 4). The mutual inductance of $L_1$ coupled to SQ1 is $M_1$. Mutual inductance of left part between terminal 3 and 4 is $kM_1$, where, $k$ is the fraction decided by location of tap; in this experimental set up, $k \approx 1/6$.

SQUID internal feedback circuit is switched between APF and SBC mode by SW2. When bias current at terminal 1 is connected to terminal 3, the SQ1 is working in APF scheme, where, $M_V = M_1$, $M_I = 0$. It is working in SBC scheme when bias current is connected to terminal 4, where, $M_V = M_1$, $M_I = kM_1$.

The test FLL was set up using an extra integrator and a feedback resistor $R_f$ connected with feedback coil $L_f$, according to Fig. 1. Parameters of SQUID and readout circuit are shown in Table 2.

The first test case (CASE I) was readout of bare SQUID. SQ1 was working as a bare SQUID in 4.2K liquid helium, where, SW2 was switched to APF mode, and $R_a$ was replaced with a 100 kΩ to avoid input coil resonance. The $\Phi_{in}$-to-$V_o$ characteristics read out with CBVA and VBCA were measured as shown in Fig. 9a and b. and current-voltage characteristics through working point ($W_1$) was shown in Fig. 9c.

Moreover, an Ultra Low Noise Amplifier (ULNA) which voltage noise is about 0.3nV/√Hz [20] was employed in the FLL to make noise performance comparisons with FLLs in CBVA and VBCA modes. The measured noise spectrums are compared in Fig. 9d. In readout of bare SQUID, flux noises measured by FLLs with CBVA and VBCA are about 7.5 μ$\Phi_0$ /√Hz, while, the flux noise measured with ULNA is reduced to 2 μ$\Phi_0$ /√Hz.

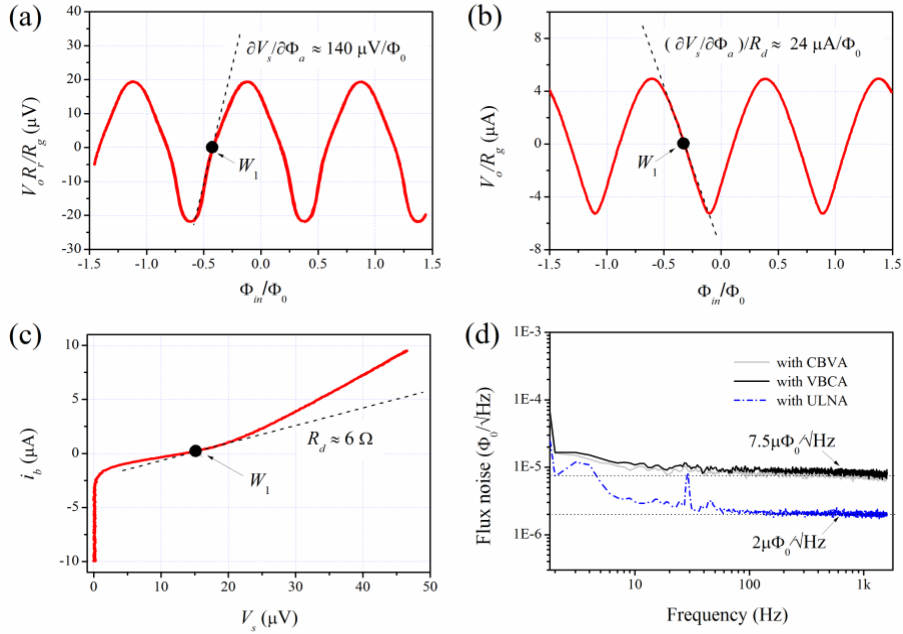

Fig. 9. (a) $\Phi_{in}$-to-$V_o$ characteristics read out with CBVA, (b) $\Phi_{in}$-to-$V_o$ characteristics read out with VBCA, (C) Current-voltage characteristic of SQ1 through the working point, and (d) the noise spectrums of FLL read out with CBVA, VBCA, and ULNA.

The second test case (CASE II) was readout of SQUID in APF mode, where, $R_a$ was set as 27Ω. The experimental results of overall $\Phi_{in}$-to-$V_o$ characteristics, current-voltage characteristic through working point, and noise spectrums read out with CBVA and VBCA were shown in Fig. 10.

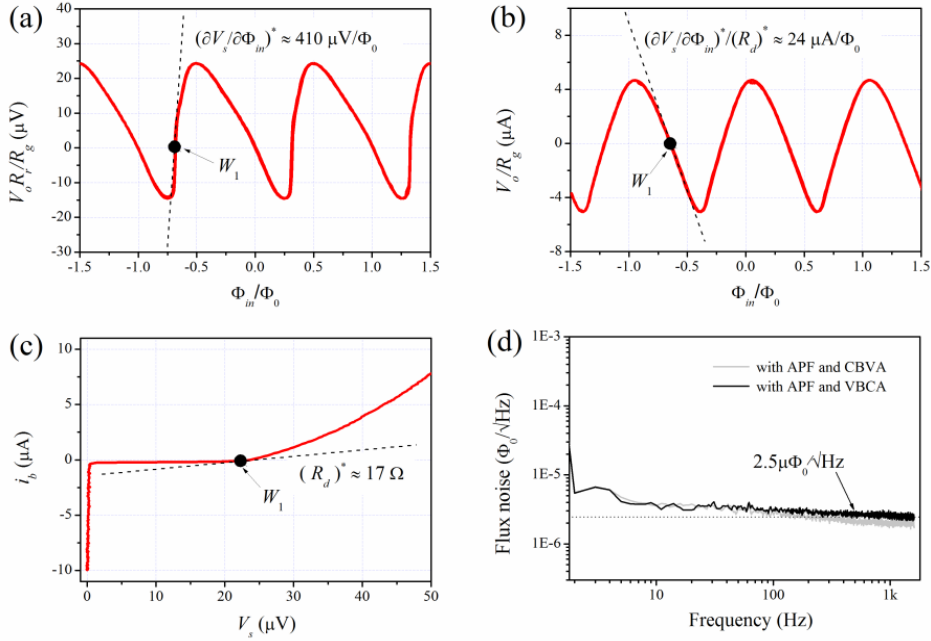

Fig. 10. (a) $\Phi_{in}$-to-$V_o$ characteristics read out with APF and CBVA, (b) $\Phi_{in}$-to-$V_o$ characteristics read out with APF and VBCA, (C) Current-voltage characteristic of SQ1 with APF through working point, and (d) The noise spectrums of FLL read out with APF in both CBVA and VBCA modes.

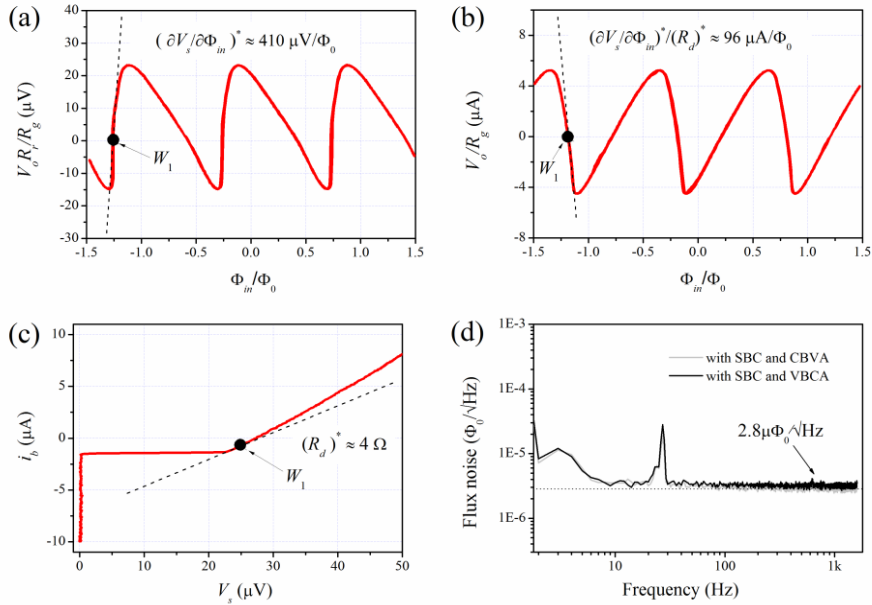

Fig. 11. (a) $\Phi_{in}$-to-$V_o$ characteristics read out with SBC and CBVA, (b) $\Phi_{in}$-to-$V_o$ characteristics read out with SBC and VBCA, (C) Current-voltage characteristic of SQ1 with SBC through working point, and (d) The noise spectrums of FLL read out with SBC in both CBVA and VBCA modes.

The third test case (CASE III) was readout of SQUID in SBC mode switched by SW2, where, $R_a$ was still 27Ω. The experimental results of overall $\Phi_{in}$-to-$V_o$ characteristics, current-voltage characteristic through working point, and noise spectrums read out with CBVA and VBCA were shown in Fig. 11.

From measured $\Phi_{in}$-to-$V_o$ characteristics in three test cases, it is proved that the VFB in APF and SBC schemes turns flux-to-voltage characteristics of SQUID measured with CBVA from symmetrical as shown in Fig. 9a to asymmetrical as shown in Fig. 10a and Fig. 11a. Meanwhile, the CFB in SBC scheme turns flux-to-current characteristics of SQUID measured with VBCA from symmetrical as shown in Fig. 9b to asymmetrical as shown in Fig. 11b.

Four key parameters on the setting working point ($W_1$) were extracted from all the experimental results as summarized in Table 3.

Table 3. Summary of parameters extracted from experimental results.

| Test case | Transfer coefficient measured with CBVA | Value ($\mu V/\Phi_0$) | Transfer coefficient measured with VBCA | Value ($\mu A/\Phi_0$) |
|---|---|---|---|---|
| I | $\partial V_s/\partial \Phi_a$ | 140 | $(\partial V_s/\partial \Phi_a)/R_d$ | 24 |
| II | $(\partial V_s/\partial \Phi_{in})^*$ | 410 | $(\partial V_s/\partial \Phi_{in})^*/(R_d)^*$ | 24 |
| III | $(\partial V_s/\partial \Phi_{in})^*$ | 410 | $(\partial V_s/\partial \Phi_{in})^*/(R_d)^*$ | 96 |
| Test case | Dynamic resistance | Value (Ω) | Equivalent flux noise | Value ($\mu\Phi_0/\sqrt{Hz}$) |
| I | $R_d$ | 6 | $\Phi_n$ | 7.5 |
| II | $(R_d)^*$ | 17 | $\Phi^*_n$ | 2.5 |
| III | $(R_d)^*$ | 4 | $\Phi^*_n$ | 2.8 |

According to (4) and (12), the flux-to-voltage transfer coefficient $\partial V_s/\partial \Phi_a$ of SQ1 and $(\partial V_s/\partial \Phi_{in})^*$ of SQ1 with APF or SBC circuit were proportional to transfer coefficient of $\Phi_{in}$-to-$V_o$ characteristics read out with CBVA. Meanwhile, the $(\partial V_s/\partial \Phi_a)/R_d$ and $(\partial V_s/\partial \Phi_a)^*/(R_p)^*$ were proportional to transfer coefficient of $\Phi_{in}$-to-$V_o$ characteristics read out with VBCA according to (5) and (13). Dynamic resistance $R_d$ of SQ1 and $(R_d)^*$ of SQ1 with APF and SBC circuit were measured from current-voltage characteristics. Finally, we picked up the equivalent white noise $\Phi_n$ with bare SQUID and $(\Phi_n)^*$ with APF and SBC scheme from noise spectrums measured in both VBCA and CBVA modes.

On the setting working point, the $\partial V_s/\partial \Phi_a$ of bare SQ1 is about 140 $\mu V/\Phi_0$, and the dynamic resistance is nearly 6 Ω. Therefore, the trans-impedance $R_V$ in both APF and SBC modes is approximately 23~25 Ω, and trans-impedance $R_I$ in SBC mode is approximately 4~5 Ω, thus, $1/(1-(R_V-R_d)/R_a) \approx 3$, and $1/(1-R_I/R_a) \approx 4$.

In the experimental results, transfer coefficient of SQUID is increased from about 140 $\mu V/\Phi_0$ to 410 $\mu V/\Phi_0$ in both APF and SBC modes; the dynamic resistance is increased from 6 Ω to nearly 17 Ω in APF mode, and decreased from 17 Ω to about 4 Ω in SBC mode. Equations (9) and (10) are verified that the flux-to-voltage transfer coefficient is modified only by VFB with $R_V$, while, the dynamic resistance is increased by VFB and decreased by CFB with $R_I$.

Meanwhile, flux noise is suppressed from 7.5 $\mu\Phi_0/\sqrt{Hz}$ to 2.5~2.8 $\mu\Phi_0/\sqrt{Hz}$ after

SQUID working in APF and SBC modes with same trans-impedance $R_V$. However, the flux noise measured with CBVA and VBCA are same in each test case, where, the bias and amplifier configurations have no effect on its noise performance.

In the first test case, the noise performance comparisons between FLL with ULNA and the one with CBVA or VBCA show that, the flux noise $\Phi_n$ is directly reduced with the reduction of amplifier noise $V_n$ as described in (6), until it is close to the intrinsic noise of SQUID.

Since, SQUID internal feedback scheme is applied only for noise suppression of preamplifier when amplifier voltage noise $V_n$ dominates the noise performance of the FLL, it will not be necessary if voltage noise of amplifier is below the SQUID intrinsic noise [21].

## 6. Conclusion

All the existing low noise direct readout schemes such as APF, NC, BCF+APF, and SBC are analyzed according to the configuration of bias mode and internal flux feedback scheme. CBVA and VBCA are two bias and amplifier modes, in which, SQUID is working as a small signal Thevenin's equivalent circuit around the working point. Transfer coefficient $\partial V_s/\partial \Phi_a$ and dynamic resistance $R_d$ of SQUID are the two key parameters deciding the open-loop gain and noise performance of direct readout circuit. They are improved by VFB and CFB circuits in internal feedback schemes.

Analyses and experimental results show that, noise performance is irrelevant to the configurations of bias and amplifier; it is only determined by flux-to-voltage transfer coefficient of SQUID. In the SQUID with internal feedback scheme, transfer coefficient is only improved by VFB with trans-impedance $R_V$, while, the dynamic resistance is increased by VFB with trans-impedance $R_V$ and decreased by CFB circuit with trans-impedance $R_I$.

Internal feedback schemes are applied only for noise suppression of amplifier which voltage noise dominates the noise performance of FLL. They will not be necessary if amplifier voltage noise is lower than SQUID intrinsic noise. Thus, finding low cost and lower noise amplifier is the ultimate solution for high performance low noise SQUID direct readout electronics.

## Acknowledgements

This work was supported by Shanghai Science and Technology Committee (Grant No: 17511103300).